%% file: main.tex
\documentclass[letterpaper, 10 pt, conference]{ieeeconf}
\IEEEoverridecommandlockouts
\usepackage{amsmath,amssymb,amsfonts}
\usepackage{algorithmic}
\usepackage{graphicx}
\usepackage{cite}
\usepackage{xcolor}
\usepackage[hyphens]{url}
\usepackage[breaklinks, colorlinks, linkcolor=purple, anchorcolor=purple, citecolor=purple, urlcolor=purple]{hyperref}
\let\labelindent\relax
\usepackage{enumitem}

\usepackage{float}
\usepackage{lipsum}

\input{shared/variables}

\newtheorem{definition}{Definition}

\begin{document}
\title{Voltage Constrained Heavy Duty Vehicle Electrification: \\Formulation and Case Study}
\author{Apurv Shukla, \IEEEmembership{Member, IEEE} Rayan El Helou, \IEEEmembership{Student Member, IEEE}, and Le Xie \IEEEmembership{Fellow, IEEE}
\thanks{}
\thanks{AS, REL, LX are with the Department of Electrical and computer Engineering at Texas A\&M University, College Station, TX e-mail\{apurv.shukla,rayanelhelou,le.xie\}@tamu.edu. This work is supported in part by NSF Grant ECCS-2038963 and U.S. Department of Energy's
Office of Energy Efficiency and Renewable Energy (EERE) under the
Solar Energy Technologies Office (SETO) Award Number DE-EE0009031.}
}

\maketitle
\thispagestyle{empty}
\pagestyle{empty}
\begin{abstract}
The electrification of heavy-duty vehicles (HDEVs) is a rapidly emerging avenue for decarbonization of energy and transportation sectors. Compared to light duty vehicles, HDEVs exhibit unique travel and charging patterns over long distances. In this paper, we formulate an analytically tractable model that considers the routing decisions for the HDEVs and their charging implications on the power grid.  Our model captures the impacts of increased vehicle electrification on the transmission grid, with particular focus on HDEVs. We jointly model transportation and power networks coupling them through the demand generated for charging requirements of HDEVs. In particular, the voltage constraint violation is explicitly accounted for in the proposed model given the signifcant amount of charging power imposed by HDEVs. We obtain optimal routing schedules and  generator dispatch satisfying mobility constraints of HDEVs while minimizing voltage violations in electric transmission network. Case study based on an IEEE 24-bus system is presented using realistic data of transit data of HDEVs. The numerical results suggest that the proposed model and algorithm effectively mitigate the voltage violation when a significant amount of HDEVs are integrated to the power transmission network. Such mitigation includes  reduction in the voltage magnitude, geographical dispersion of voltage violations and worst-case voltage violations at critical nodes. 
\end{abstract}

\section{INTRODUCTION}
\input{paper/introduction}
\section{MODEL}
\input{paper/formulation}

\section{CASE STUDY}
In this section, we present a set of experiments investigating the impact of realistic HDEV transportation patterns on a transmission grid under a set of alternative fleet dispatch mechanims. We focus comparison on two types of scenarios: (a) one where we adopt a grid-agnostic HDEV spatiotemporal scheduling mechanism to determine where and when charging occurs, and (b) one where we adopt the solution to the proposed joint ACOPF-HDEV co-optimization presented in \eqref{eqn:formulation-combined}. Our results indicate that under the co-optimization, there is a clear improvement in the voltage profile while maintaining required business travel requirements on heavy-duty vehicles.

The simulation is performed as follows. First, we consider a simplified model of the Texas grid by using an IEEE 24-bus \cite{ordoudis2016updated} transmission grid model scaled to meet realistic Texas-based generation and demand criteria. Second, we consider between 30,000 to 50,000 HDEVs present in the network, which a suitable baseline for HDEV impact in Texas according to reference \cite{el2022impact}, and they are each constrained by having to meet some business travel criteria. Travel criteria is defined in the context of this work as a set of required arrival and departure times, initial and final levels of charge, and initial and final physical locations.

For the remainder of this case study, we reference the following example for illustration purposes. Consider some fleet $h\in\mathcal{H}$ with the following travel criteria:
\begin{align*}
    \overline{\psi}^{h,\texttt{in}}_v \gets N^h &\quad \text{for} \quad v = (\text{Bus 1, 50\% Charge, Hour 1}) \nonumber \\
    \overline{\psi}^{h,\texttt{out}}_v \gets N^h &\quad \text{for} \quad v = (\text{Bus 22, 25\% Charge, Hour 7}) \nonumber
\end{align*}

That is, this fleet of vehicles begins its business trip at hour 1 and the location corresponding to bus 1, with 50\% of battery charged. It is required to arrive at no later than hour 7 at the location corresponding to bus 22 with a minimum battery charge level of 25\%.

\begin{figure*}[!htbp]
\centering
\includegraphics[width=0.7\textwidth, trim={100, 100, 100, 100}]{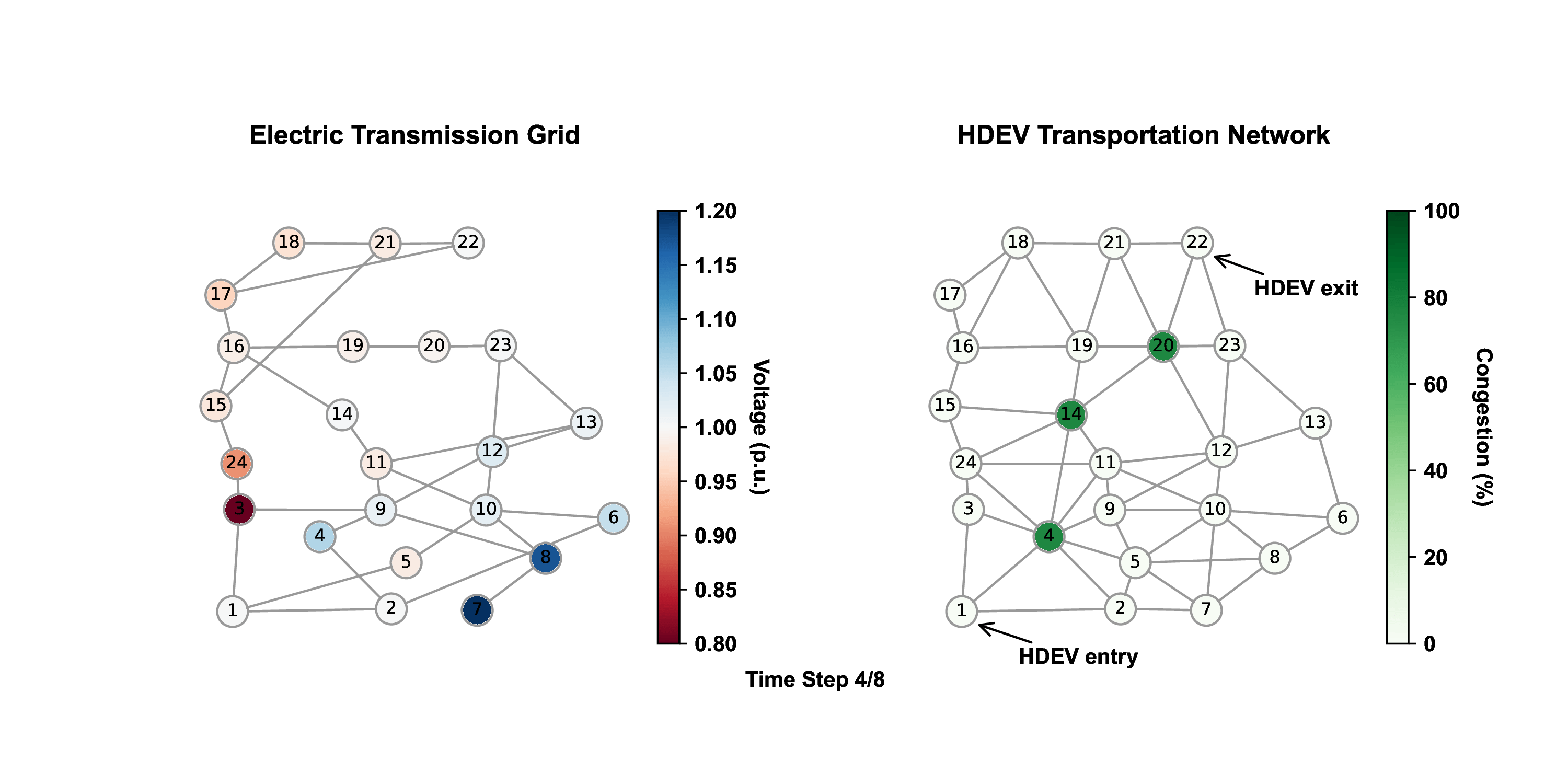}
\caption{Voltage distribution without co-optimization.}
\label{fig:map_before}
\end{figure*}

With this set of business requirements on the fleet of vehicles, we compare two spatiotemporal HDEV dispatch approaches to study the impact on the system-wide transmission-level voltage:
\begin{enumerate}
    \item The vehicles follow the shortest path and stop to charge when they run out of energy. This yields a path of buses ($1\to4\to14\to20\to22$). See Fig. (\ref{fig:map_before}) for the numbers of buses in the IEEE 24-bus case.
    \item Solve the joint ACOPF-HDEV co-optimization, which yields a spatiotemporal redistribution where vehicles strategically time their charging to avoid congestion at key stations and key hours.
\end{enumerate}

\begin{figure}[!htbp]
\centering
\includegraphics[width=\linewidth]{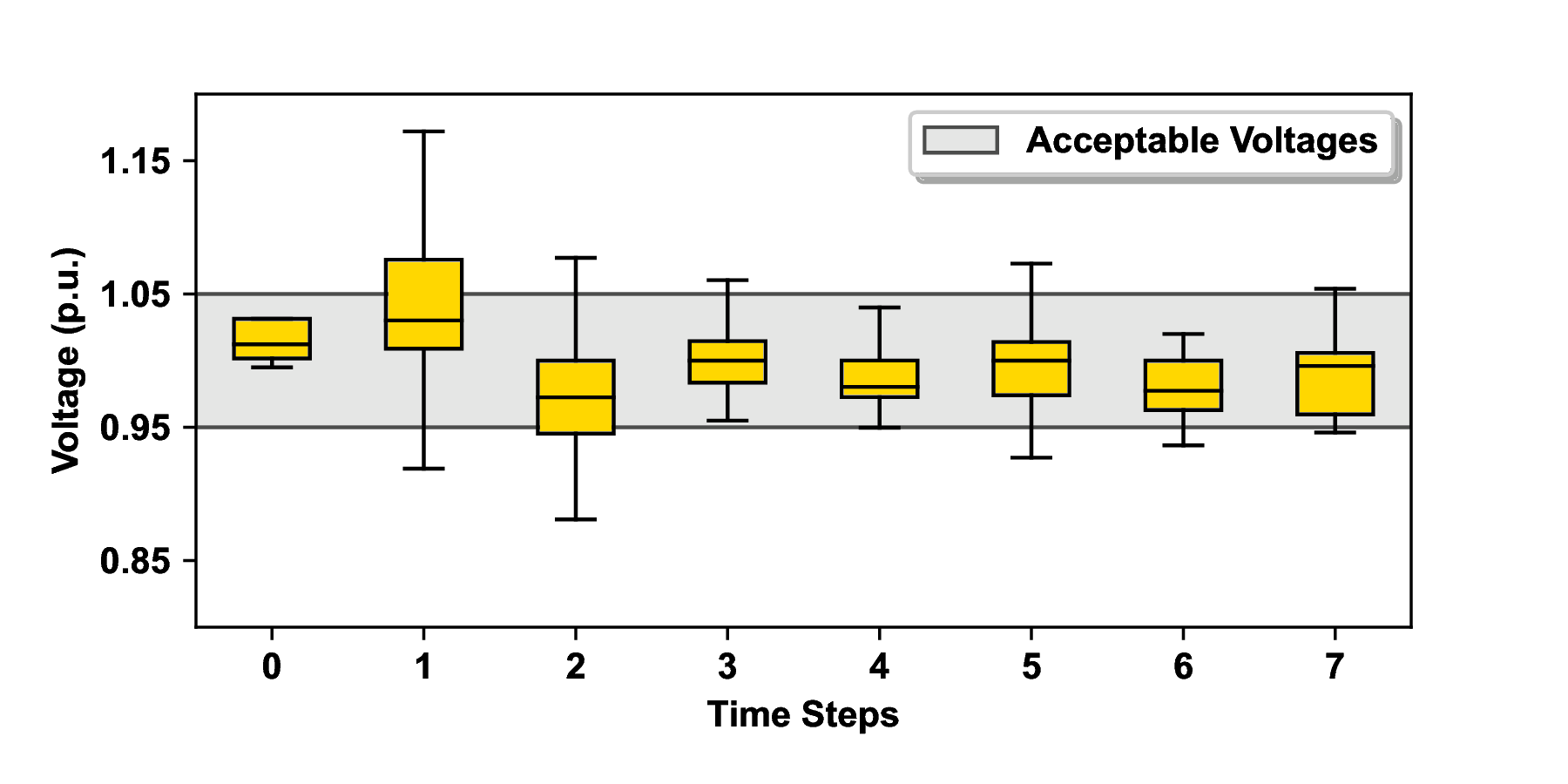}
\caption{Voltage distribution without co-optimization.}
\label{fig:voltages_before}
\end{figure}

\begin{figure}[!htbp]
\centering
\includegraphics[width=\linewidth]{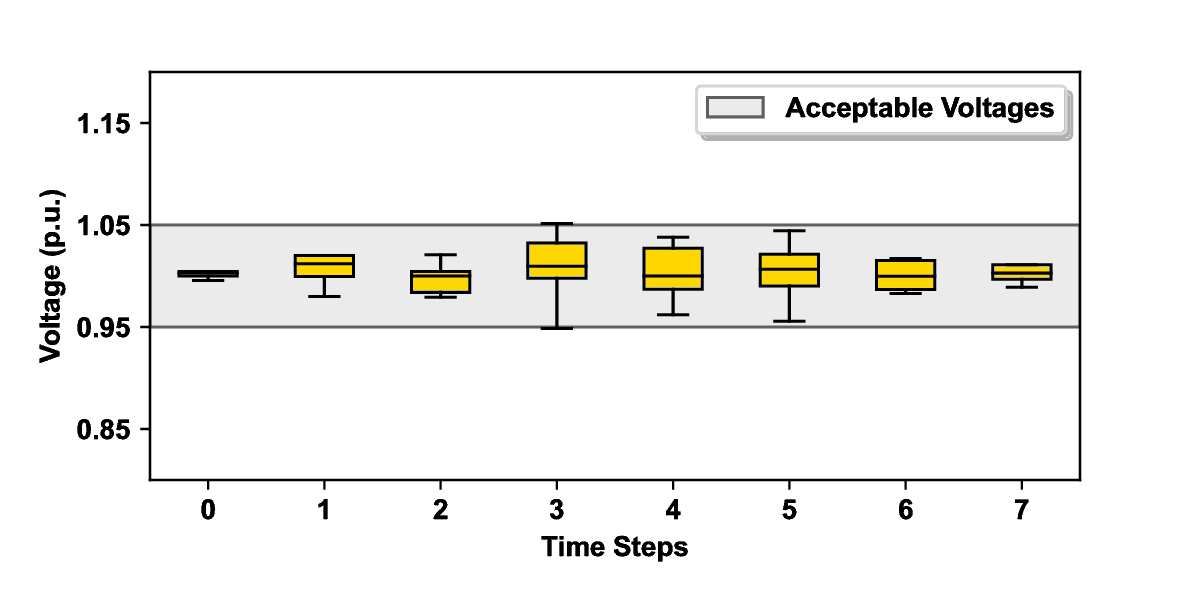}
\caption{Voltage distribution with $\ell_{\infty}$ co-optimization.}
\label{fig:voltages_after_Linf}
\end{figure}

\begin{figure}[!htbp]
\centering
\includegraphics[width=\linewidth]{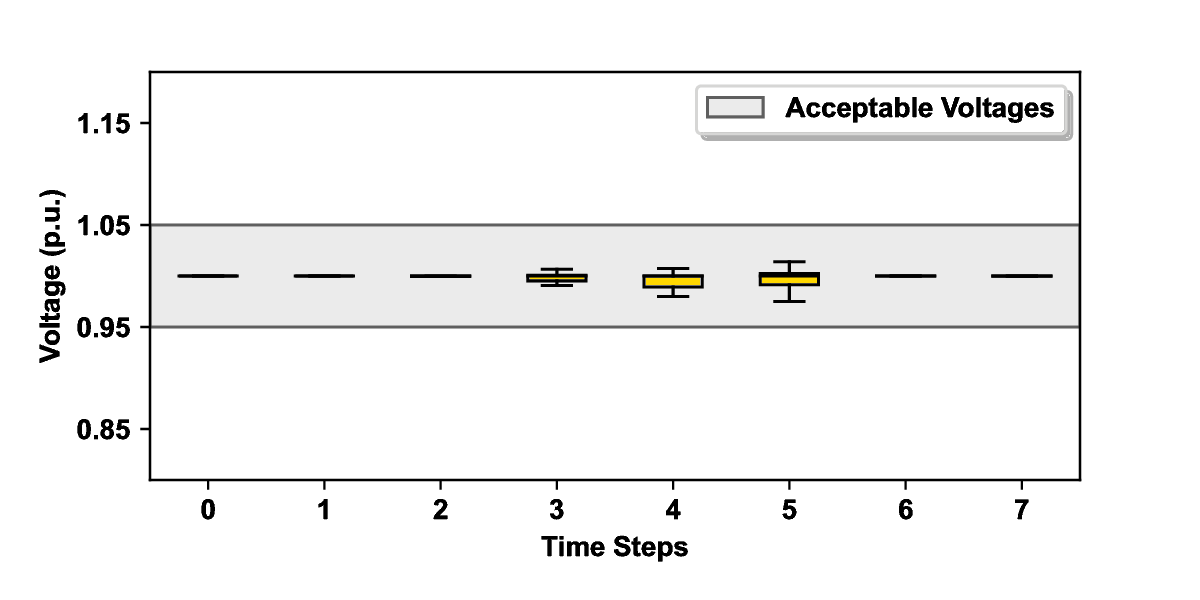}
\caption{Voltage distribution with $\ell_{1}$ co-optimization.}
\label{fig:voltages_after_L1}
\end{figure}

The results for voltage violations are shown in Fig. (\ref{fig:voltages_before}, \ref{fig:voltages_after_Linf}, \ref{fig:voltages_after_L1}), illustrating that the first approach yields many voltage violations across time unlike the proposed approach which only yields much less voltage deviations. To illustrate acceptable voltage ranges, we shade the region between 0.95 and 1.05 per unit voltage in either figure, which is typically acceptable in a power systems context. While Fig. (\ref{fig:voltages_after_Linf}, \ref{fig:voltages_after_L1}) are both based on the proposed approach, the former is based on the $\ell_{\infty}$ norm objective function for voltage violations and the latter is based on the $\ell_{1}$ norm.

As shown in Fig. (\ref{fig:voltages_before}), under the grid-agnostic HDEV spatiotemporal distribution of vehicles, the voltage violations significantly exceed the acceptable band of voltages. In contrast, under both $\ell_{\infty}$ and $\ell_{1}$ norms using the proposed co-optimization, voltage is kept within the acceptable band. The results in Fig. (\ref{fig:voltages_after_Linf}, \ref{fig:voltages_after_L1}) suggest that perhaps the $\ell_{1}$ norm is better for voltage regulation purposes. However, this could also yield an increase in usage of physical resources to provide this regulation. A deeper exploration of the optimal use of such resources is beyond the scope of this work. In this paper, we simply assume that all resources are constrained by upper and lower limits on active and reactive power and a cost function, as shown in formulation (\ref{eqn:formulation-combined}).

\begin{figure*}[!htbp]
\centering
\includegraphics[width=0.7\textwidth, trim={100, 100, 100, 100}]{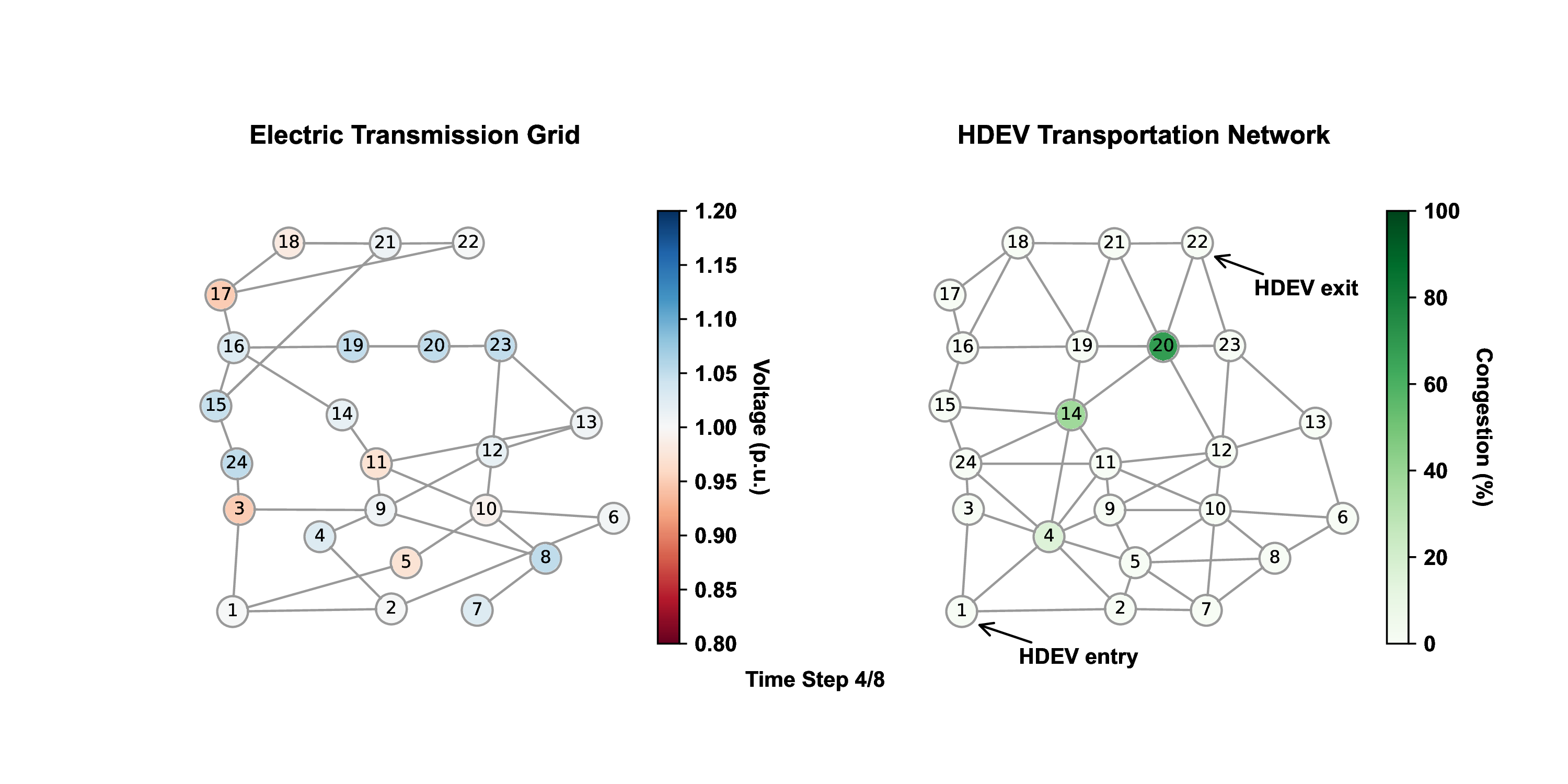}
\caption{Voltage distribution with $\ell_{\infty}$ co-optimization.}
\label{fig:map_after_Linf}
\end{figure*}

\begin{figure*}[!htbp]
\centering
\includegraphics[width=0.7\textwidth, trim={50, 50, 50, 50}]{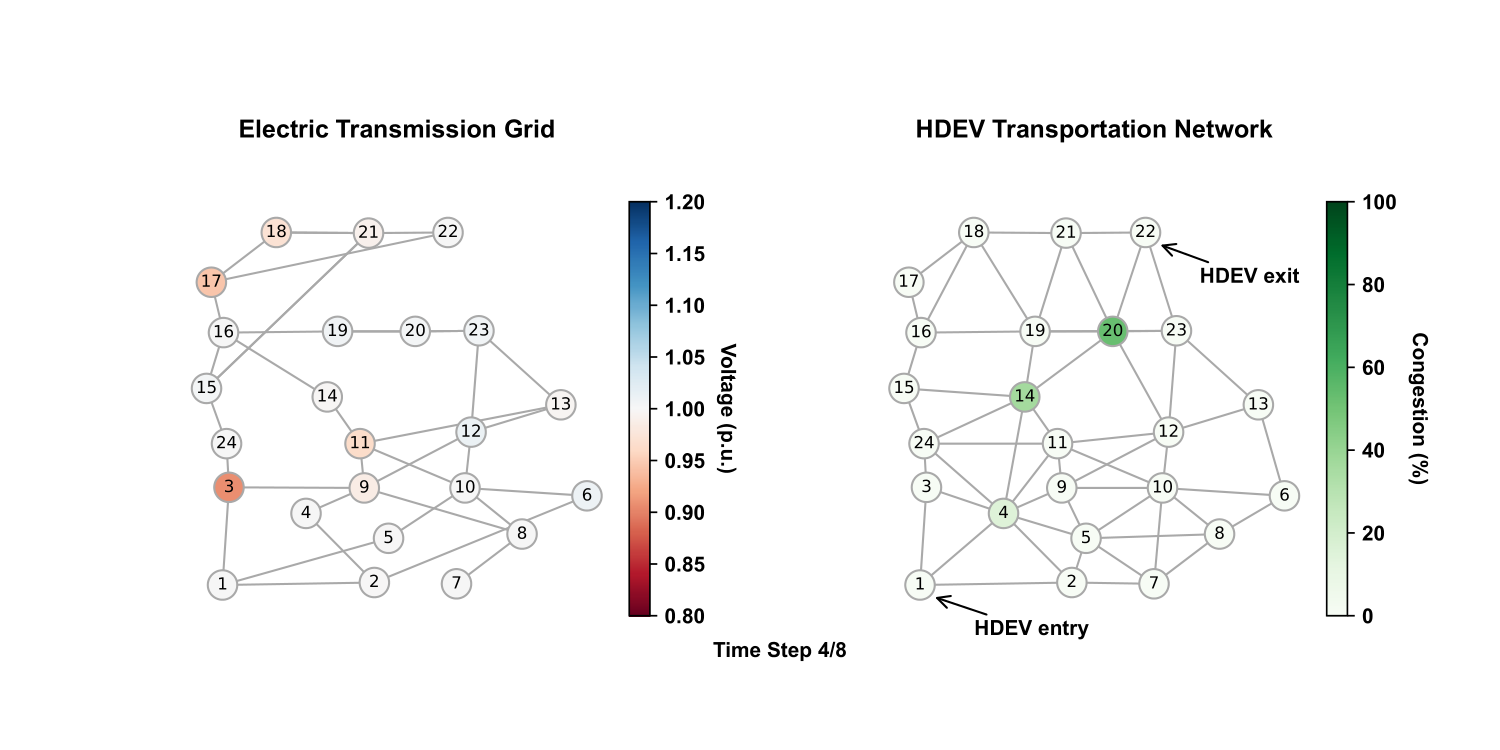}
\caption{Voltage distribution with $\ell_{1}$ co-optimization.}
\label{fig:map_after_L1}
\end{figure*}

To further illustrate the difference between the grid-agnostic approach and the proposed approach, we present a visual representation of both the electric transmission grid and the HDEV transportation network at a certain snapshot in time, specifically showing a spatial depiction of the condition of the system in the simulation for any of the cases. The results are shown in Fig. (\ref{fig:map_before}, \ref{fig:map_after_Linf}, \ref{fig:map_after_L1}). From these results, we can observe that the proposed approach manages to spatially dilute the voltage deviations across the electric transmission grid (figures on left side), mainly by strategically changing the following on the HDEV transportation side (figures on right side):
\begin{itemize}
    \item location to assign most charging congestion
    \item time to wait before charging
\end{itemize}
Note that we do not explicitly state how to spatiotemporal redistribute the charging, but the proposed co-optimization automatically yields such results. We now direct the readers attention to Fig (\ref{fig:map_after_Linf}, \ref{fig:map_after_L1}) specifically. Indeed, both results clearly out-perform the grid-agnostic approach (Fig. (\ref{fig:map_before})) voltage-wise. However, when using the $\ell_{\infty}$ norm as opposed to the $\ell_{1}$ norm for voltage deviations in the objective function, i.e. Fig. (\ref{fig:map_after_Linf}) as opposed to , Fig. (\ref{fig:map_after_L1}) respectively, \textbf{more number of buses} incur voltage violations, albeit small in magnitude. In contrast, the $\ell_{1}$ norm yields voltage violations which are \textbf{more at the worst buses} yet less number of buses. This kind of distinction between the two norms is expected as described in the formulation section, and we leave it to a power systems operator to judge which is more suitable for their grid management application.

Nonetheless, for either norm, as shown on the right sides of Fig. (\ref{fig:map_after_Linf}, \ref{fig:map_after_L1}), there is clearly less congestion per charging station than in the case without co-optimization. These results suggest that with co-optimization of HDEV spatiotemporal scheduling and optimal power flow, business travel criteria can still be met while yielding significantly improved voltage profiles in the electric transmission grid.

\section{CONCLUSION AND FUTURE WORK}
In this paper, we propose a first step towards modeling the routing and scheduling of heavy duty vehicles with particular focus on their impact on the power grid. Our simulation results corroborate that adding penalty functions to the joint transportation and power flow problem does indeed result in more even geographic distribution of HDEVs and helps us in minimizing voltage violations from nominal levels. As expected, different penalty measures serve different purpose but all of them perform better than separately optimizing transportation and energy networks.
\medskip

In the future we would like to pursue several directions pertaining to model and algorithm development. First, our model can be expanded to take into account distribution network. On the algorithmic side we would like to develop efficient algorithms for solving the co-optimization problem involving both energy and transportation networks. Finally, reformulations and efficient practical deployment of the proposed model in real-life systems remains a challenge.

\clearpage

\bibliographystyle{IEEEtran}
\bibliography{ref}

\section*{Appendices}
\input{paper/ACOPF}

\end{document}

%% file: shared/variables.tex
\def\cA{\mathcal{A}}

\def\cE{\mathcal{E}}

\def\cG{\mathcal{G}}
\def\cH{\mathcal{H}}
\def\cI{\mathcal{I}}

\def\cL{\mathcal{L}}

\def\cN{\mathcal{N}}
\def\cO{\mathcal{O}}

\def\cV{\mathcal{V}}

\newcounter{problem}
\newcounter{save@equation}
\newcounter{save@problem}

\makeatletter
\newenvironment{problem}
 {\setcounter{problem}{\value{save@problem}}%
  \setcounter{save@equation}{\value{equation}}%
  \let\c@equation\c@problem
  \renewcommand{\theequation}{P\arabic{equation}}%
  \subequations
  }
 {\endsubequations
  \setcounter{save@problem}{\value{equation}}%
  \setcounter{equation}{\value{save@equation}}%
 }

%% file: paper/introduction.tex
Recent surge of market penetration of electric vehicles (EVs), calls for development of scalable design and operation of charging infrastructure in support of the decarbonization of the mobility sector.  Of 30\%  U.S. total green-house gas emissions~\cite{greenhouse}, $25\%$ can be attributed to medium-duty and heavy-duty vehicles (HDVs)~\cite{climateenergy}, i.e., 7.5\% of the total emissions are due to HDVC. Of particular interest is the potential impact of the electrification of HDVs, specifically Class 7, 8 and 9 vehicles exceeding $26,000$ pounds~\cite{annualvehicle}. The state of Texas has $7\%$ of the nationwide total of $3.91$ million Class 8 heavy vehicles \cite{federalhighway} of the approximately $275,000$ HDVs registered. Each HDV travels on average over $62,000$ miles annually - nearly $5.5$ times the distance traveled by a typical passenger car~\cite{annualvehicle}. The significance of electrification is further amplified by noting the average heavy-duty electric vehicle (HDEV) is expected to draw anywhere between 75 kW and 600 kW while charging~\cite{unterlohner2020comparison}. Such fundamental shifts in electrification require a substantial restructuring and shift from fuel-based to electricity-based charging
infrastructure. This would lead to significant disruptions for both electric grid operators and external market participants. This makes it urgent and imperative for policy makers to adopt a data-driven, scientific approach to reconcile the market adoption of HDEVs with the realities of existing grid infrastructure 
There has been a burgeoning interest in studying recent the impact of HDEV on the power grid. \cite{el2022impact} demonstrate the impact of HDEV charging decisions on the power grid.\cite{prussi2022comparing} study how different types of fuels effect decarbonization of heavy-duty vehicles.

From the perspective of an electric or transportation network operator two critical concerns arise given the surge in EV penetration: (i) mismatch of demand allocation and infrastructure capacity, leading to inefficient use of charging stations and (ii) mismatch of demand and supply in the power grid, leading to voltage instability, and line capacity violations~\cite{tushar2015cost}. Existing literature on these issues can be divided into three categories:

\begin{enumerate}[label = (\alph*)]
\item The first category aims to manage the effects of EV charging load on the power grid, at both the distribution and transmission levels, with most previous work capturing plug in decisions at specific charging stations as exogenous random processes. Some recent works model the effect of EV mobility patterns on the spatio-temporal distribution of charging load~\cite{mu2014spatial,de2014gis}. Recent work~\cite{alizadeh2018retail} focuses on designing mechanisms for fleet deployment and charging at optimal times and locations. 
\item Other work focuses on the scheduling problem of individual EV fleets considering the time needed to charge,
limited energy stored in the battery, and range anxiety~\cite{goeke2015routing,wang2015optimal,artmeier2010shortest} assuming that the electricity prices are given. The potential of an EV fleet for vehicle-to-grid (V2G) services considering their mobility patterns is studied in~\cite{yu2015balancing}.
\item  The third category considers the interaction between power and transportation networks. For EVs that are parked while charging, \cite{khodayar2013electric} consider a case where the operator tracks the mobility of large shared fleets of EVs and their energy consumption and designs optimal multi-period vehicle-to-grid (V2G) strategies. For individual EVs (i.e., those not belonging to a fleet)~\cite{alizadeh2016optimal} consider mobility-aware EV demand management problem to calculate wholesale market clearing prices.~\cite{he2013optimal} study the optimal deployment of EV public charging stations by formulating it as a mathematical program with equilibrium constraints. 
\end{enumerate}
\medskip 
On top of the above challenges, HDEVs have several specific challenges unaddressed in antecedent literature:
i) the threat of grid instability due to line fluctuations and voltage deviation from nominal levels ii) sub-optimal design of routing and scheduling policies that do not account for vehicle charging levels especially since HDEVs are on road 80\% of time. The purpose of this paper is to manage the effects of HDEV movement and charging on the power grid, at transmission levels. 

\begin{figure*}[!htbp]
\centering
\includegraphics[width=\textwidth]{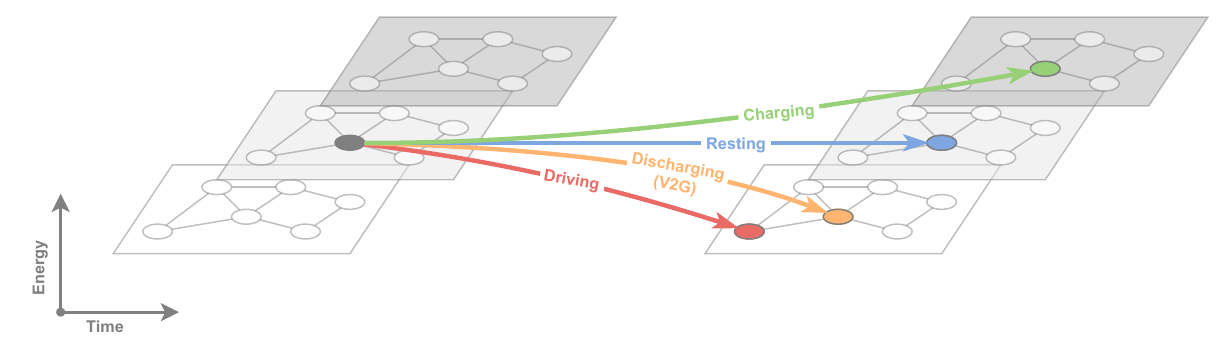}
\caption{Time-Energy Expanded Graph: Illustrations of movements along time and energy dimensions}
\label{fig:expanded_graph}
\end{figure*} 

\medskip
In this paper, we consider a dynamic energy and transportation network to capture variations in travel times, routing requests, and energy requirements from the perspective of an Independent System Operator (ISO) who is interested in suggesting a optimal route and energy consumption of a fleet of HDEVs. To do so, we consider the interaction of ISO with individual EV drivers and propose a model that \textit{simultaneously} prescribes the optimal route (subject to the driver's constraints) as well as the optimal power dispatch schedule for the ISO. We do not focus on pricing strategies and equilibrium considerations that arise from those. Our first primary contribution is to propose a tractable and computationally efficient model of the transportation and energy networks associated with scheduling HDEVs. Existing models are either intractable due to combinatorial explosion in the number of variables or the associated optimization problem are non-convex. Further, existing models completely eschew aspects of HDEV scheduling pertinent for power and transportation network operation. In contrast, our model uses significantly smaller number of variables and the associated optimization problem is convex. Further, in line with our main goal of analyzing the impact of HDEV integration on the reliability of transmission and distribution grids the model minimizes grid voltage violations while determining optimal routes for HDEV scheduling. Our second contribution is the development of a Python-based simulation tool, to study the consequences of HDEV electrification, exposing the limitations imposed by existing grid infrastructure. The parameters used to realistically model different components of the simulation (such as wind~\cite{freeman2018system}, solar~\cite{sengupta2018national}, vehicle data~\cite{walkowicz2014fleet} and electric grid simulation~\cite{thayer2020easy}). We utilize an IEEE-24 bus system that realistically represents multiple distribution grids in Travis county, Texas~\cite{birchfieldcombined}, and transit data pertaining to HDEVs. Our simulation studies demonstrate the advantages of co-optimizing our power flow and transportation objectives as compared to a decoupled approach on grid reliability, congestion on the transportation network and charging decisions of the HDEV. To the best of our knowledge, this is the first study on the effects of HDEVs on a combined energy and transportation network.

%% file: paper/formulation.tex
Our main goal in this section is to propose a dynamic energy-transportation network model that captures scheduling and routing requirements for a fleet of HDEVs. A transportation network can be modeled as a graph $\cG \left(\cV,\cE\right)$, where $\cV$ denote the set of vertices and $\cE$ denote the set of edges. To model the movement of vehicles within the state of Texas, we use such a graph where nodes represents the charging station locations and edges represent the roads along which any fleet of HDEVs must travel to relocate from one node to another. Edges are weighted by the average distance required to traverse from either node to the other (where each edge connects two nodes). Typically, vehicles would use a path on the transportation network that can take them from source to destination, obtaining such a network using a shortest path problem, where each arc has a weight equal to the travelling time on that arc. We now extend this transportation network to incorporate energy levels and time. 
\smallskip

A heavy-duty electric vehicle runs on electric storage, where each vehicle is associated with a state of charge (SOC), which decreases only when the vehicle moves, and increases only when the vehicle is charging. In contrast to smaller electric vehicles, HDEVs not only require larger amounts of energy (kWh) to fully charge, but also demand more charging power (kW) to reasonably serve the vehicles. To model these energy requirements on top of a transportation network, we expand a vertex in the transportation graph to include (discrete) energy levels $e \in [e_{\min},\ldots,e_{\max}]:= \cE $. Finally, in order to consider discrete time steps and the planning horizon $T$, we add a third dimension representing time to each vertex. This leads us to define an extended energy-transportation network in Definition~\ref{defn:energy-transportation}.

\begin{definition}[Extended Energy-Transportation Network]
\label{defn:energy-transportation}
Let $\cG^{e}=(\cN,\cA)$ denote a directed graph where each vertex $v \in \cV$ is a $3$-tuple $(i,e,t) \in \cN \subseteq [\vert \cV \vert] \times [\vert \cE \vert] \times [T]$ with the first component representing the physical location, the second representing the energy levels and the third time. An arc $a \in \cA$ connects tail node $(i,e,t)$ and head node $(i',e',t')$ iff: 
\begin{enumerate}
\item There is no movement backward in time, i.e., $t' > t$
\item Charging Arc: Increase in energy level and same physical location implying $e < e_{\max}$ and $e' = e+1, \ i'=i$.
\item V2G Discharging Arc: Decrease in energy level and same physical location, i.e., V2G discharging implying $e > e_{\min}$ and $e' = e-1$ and $i' = i$.
\item Driving arc: $e > e_{min}, \ e' = e-1, \ i' =\text{neighbor}(i)$, where $\text{neighbor}(i)$ represents 
\item Resting arc: $e' = e$ and $i' = i$
\end{enumerate}
\end{definition}

We consider the problem of managing a given fleet of HDEVs. The purpose of fleet management is to ensure that there is no congestion in the transportation graph $\cG$ and the expanded energy-transportation $\cG^{e}$. In the subsequent section, we will suppress the dependence of the constraint and equation on $t$ which implies that these relationships need to hold for all $t \in [T]$.

\subsection{Fleet Management System}
We begin by considering a fleet $\cH$ of HDEVs. The fleet management system also needs to prescribe routes based on the fleet's requirements. Let $\cO_{a}$ denote the set of arcs orginating at node $v$ and $\cI_{v}$ denote the incoming arcs at node $v$.Further, $\psi^{h,\text{in}}_{v}$ is the exogenous injection of traffic of fleet $h$ at node $v$ and $\psi^{h,\text{out}}_{v}$ denotes the exogeneous disappearence of the HDEV from the network. This process is subject to the following constraints: 
\begin{eqnarray}
 & \overline{\psi}^{h,\texttt{in}}, \overline{\psi}^{h,\texttt{out}} \sim \text{Travel Criteria Distribution} \nonumber \\
& \text{s.t.}~ \sum_{v\in \mathcal{V}} \overline{\psi}^{h,\texttt{in}}_v = \sum_{v\in \mathcal{V}} \overline{\psi}^{h,\texttt{out}}_v \nonumber \\
&\text{s.t.}~~ \overline{\psi}^{h,\texttt{in}}_v, \overline{\psi}^{h,\texttt{out}}_v \geq 0 \nonumber \\
& \sum_{u\in\mathcal{U}_v} \psi^{h,\texttt{out}}_u = \overline{\psi}^{h,\texttt{out}}_v \label{eqn:node-balance-2} \nonumber \\
& \psi^{h,\texttt{out}}_v \geq 0 \nonumber 
\end{eqnarray}
where, $\bar{\psi}^{h,\text{in}}_{v}$ and $\bar{\psi}^{h,\text{out}}_{v}$ is the system operator specified injection/rejection from the system. Let $\lambda_{a}(t)$ denote the flow of traffic on arc $a \in \cA$ in the extended energy-transportation network. The flow of HDEV $h \in \cH$ satsifes the following flow-balance constraints at all times $t \in [T]$:
\begin{eqnarray*}
& \sum_{a\in\mathcal{O}_v} \lambda_a^h - \sum_{a\in\mathcal{I}_v} \lambda_a^h = \overline{\psi}^{h,\texttt{in}}_v - \psi^{h,\texttt{out}}_v \label{eqn:node-balance-1} 
\end{eqnarray*}

\subsection{Power Dispatch Problem}
\label{sec:framework}
Each HDEV has direct impact on the electric power grid only when it is charging at an electric vehicle charging station (EVCS). While there may be several EVCS’s connected to a single distribution grid, we assume that no HDEV charges at more than one EVCS within the same distribution grid in the short run, since HDVs travel long dis tances after each charge. Thus, when modelling the impact of HDEVs
on distribution grids, it suffices to capture only their local power consumption, without their movement within the neighborhood itself. Conversely, since EVCS’s are geographically dispersed across multiple locations in the transmission grid (state-wide or wider), we believe that it is necessary to consider the movement of a fleet of vehicles within each transmission grid’s encompassing geographic region to capture heir spatio-temporal impact on critical grid parameters such as grid voltages. 
Given the demand generated by considering the mobility in the transportation network, we determine the optimal power flow in the power network using the AC optimal power flow (AC-OPF)~\cite{bienstock2019strong,roald2017chance}. The AC-OPF problem is a nonlinear, nonconvex problem that models the power flow in electrical networks, determining nodal voltages levels, angles and line flows~\cite{lavaei2011zero,molzahn2019survey,kocuk2018matrix,chen2017spatial}. From the transportation graph $\cG$, we assume that every physical location $\cV$ also represents a bus and the edges $\cE$ represent a power line. We assume that there is one conventional generator with active and reactive power outputs ${p}_{G,i}, q_{G,i}$, one demand $p_{D,i}, q_{D,i}$ per node that is influenced due to the presence of HDEVs. Nodes without generation or load can be handled by setting the respective entries to zero, and nodes with multiple entries can be handled through a summation. To model generation and control at different types of buses, we add subscripts $P Q, P V$, and $\theta V$ to distinguish between $PQ, PV$ and $\theta V$ (reference) buses. We model the problem of determining optimal power flow in the energy network, using AC-OPF. Therefore, every node has the following associated constraints:
\begin{enumerate}
\item Nodal Constraints: At every node $i \in \cV$, there are three constraints: 
\begin{enumerate}
\item Nodal demand balance which ensures both active, $P_i$ and reactive power $Q_i$ are balanced. 
\begin{eqnarray*}
P_i^d &=& \overline{P}_i^d + x_i  \\
Q_i^d &=& \overline{Q}_i^d + \alpha^{PQ}x_i 
\end{eqnarray*}
In this formulation, $x_i$ denotes the excess demand due to the presence of HDEVs at node $i$. A fraction of this demand $\alpha^{PQ}$ of the real demand is incurred as reactive power.
\item Nodal flow balance:
Constraints representing that the total throughput at a node is equal to the net generation and demand at that node. 
\item Generation Limits: The generation limits reflect the minimum and maximum capacity of generation at individual nodes.
\begin{eqnarray*}
\underline{P}_i^g \leq P_i^g \leq \overline{P}_i^g \\
\underline{Q}_i^g \leq Q_i^g \leq \overline{Q}_i^g 
\end{eqnarray*}
\end{enumerate}
\item Line Flow: For every line $ij$, let $\theta_{ij}$ denote the difference in phase angles between nodes $i$ and $j$ and $y_{ij}$ denote its admittance. We add the following thermal limit to line constraints: 
\[
\vert y_{ij}\vert^2\left[\vert V_i\vert^2 + \vert V_j\vert^2 - 2\vert V_i\vert \vert V_j \vert \cos{(\theta_{ij})}\right] \leq \overline{I}_{i,j}^2
\]
\end{enumerate}
The AC-OPF formulation mentioned before is non-linear and non-convex and is computationally inefficient without further assumptions. In order to make our formulation tractable, we propose an analytical reformulation of the AC-OPF problem by utilizing a first-order Taylor approximation of the branch-flow and thermal-limit constraints about an operating point. For this purpose, let $(\mathbf{\theta}_{0},\mathbf{V}_{0}, \mathbf{P}_{0}, \mathbf{Q}_{0})$ denote the vector of operating point and let $\mathbf{J}^{P}_{V}$ denote the Jacobian of $\mathbf{P}$ wrt $\mathbf{V}$. For example, for the branch flow constraint, this becomes:
\begin{align*}
\mathbf{P^g} - \mathbf{P^d} = \mathbf{P^g}_0 - \mathbf{P^d}_0 + \mathbf{J}^P_V(\mathbf{V}-\mathbf{V}_0) + \mathbf{J}^P_\theta(\mathbf{\theta}-\mathbf{\theta}_0)\\
\mathbf{Q^g} - \mathbf{Q^d} = \mathbf{Q^g}_0 - \mathbf{Q^d}_0 + \mathbf{J}^Q_V(\mathbf{V}-\mathbf{V}_0) + \mathbf{J}^Q_\theta(\mathbf{\theta}-\mathbf{\theta}_0)
\end{align*}
Other constraints can be linearized similarly. The final ingredient in our formulation is the objective function that penalizes the deviation of the nodal voltage from nominal operating point. To this end, we propose the following objective: 
\[
\textbf{cost}(\mathbf{P}^{g}(t))  = \sum_{i \in \cV} c_{1,g}P^{2}_{i,g} + c_{1,i} P_{i,g} + c_{0,i} + \Phi(V_i)
\]
where, $\Phi(\cdot)$ denotes a penalty term on the voltage magnitude. Of particular interest to us are the following penalties:
\begin{enumerate}
\item $\ell_{2}$ co-optimization: $\Phi(V_{i}) := \sum_{i \in \cV} \left(V_i - \bar{V}_i \right)^{2}$: This penalty term ensures that deviation from operation point $\bar{V}_i$ will be minimized. 
\item $\ell_{1}$ co-optimization:
$=\Phi(V_{i}) := \sum_{i \in \cV} \vert V_i - \bar{V}_i \vert$: In addition to ensuring that deviations are minimized, this penalty ensures a sparse geographical distribution of voltage deviations enhancing grid reliability.
\item $\ell_{\infty}$ co-optimization: $\Phi(V_{i}) := \max_{i \in \cL} \vert V_i - \bar{V}_i \vert$: This penalty focuses on minimizing the largest $\vert \cL  \vert$ nodal voltages where the set $\cL$ is the set of top-$L$ largest nodal voltages $\bar{V}_{i}$.
\end{enumerate}

Several remarks are in order regarding our modeling choices. First, it is widely established that DC-OPF is the standard for computing power flows in practice. However, given our motivation to study the impact of placement of HDEVs on voltage deviation in the grid, DC-OPF doesn't suit our purposes since it doesn't take into account nodal voltages. Secondly, among the several possible linearization schemes, our proposed reformulation includes the complete set of non-convex AC power flow equations for the specified operating point. Crucial for our purposes, is to ensure feasibility of the power flow problem about an operating point which is not guaranteed by approximations, partial linearization or relaxations of the problem. Finally, the choice of the penalty function is critical in ensuring reliability of the network. Our combined model (transportation with energy network) is presented in Appendix~\ref{appendix:linear-model}. 


%% file: paper/ACOPF.tex
\newcommand{\Graph}{\mathcal{G}}
\newcommand{\Vertices}{\mathcal{V}}
\newcommand{\Arcs}{\mathcal{A}}
\subsection{Combined Transportation and Power Problem}
\label{appendix:linear-model}
The complete formulation is given in the next page. 
\begin{table*}[h]
\begin{problem}
\label{eqn:formulation-combined}
\begin{align}
\underset{\mathbf{P^g},\mathbf{\lambda}}{\text{minimize}} \quad & \sum_{t=1}^{T}\textbf{cost}(\mathbf{P^g}(t)) + \Phi(V(t)) \\
\text{s.t.} \quad & \forall t \in [1,T]: \nonumber\\
\textbf{Power Grid Constraints}:\quad & \nonumber\\
& \theta_\text{slack} = 0 && \text{(reference angle)} \\
& V_i = \overline{V_i} \quad \forall i\in\mathcal{N}^\text{AVR} && {\begin{aligned} & (\text{automatic} \\ & ~~~\text{voltage regulation}) \end{aligned}} \\
\forall i\in\mathcal{N}&: \nonumber\\
& P_i^d = \overline{P}_i^d + x_i \quad && \text{(bus demand)} \\
& Q_i^d = \overline{Q}_i^d + \alpha^{PQ}x_i \quad && \text{(bus demand)} \\
& P_i^g - P_i^d = \sum_{j\in\mathcal{N}} V_iV_j\left[G_{ij}\cos(\theta_i - \theta_j)
+ B_{ij}\sin(\theta_i - \theta_j)\right] && \text{(flow balance)} \\
& Q_i^g - Q_i^d = \sum_{j\in\mathcal{N}} V_iV_j\left[G_{ij}\sin(\theta_i - \theta_j)
- B_{ij}\cos(\theta_i - \theta_j)\right] && \text{(flow balance)}\\
& \underline{P}_i^g \leq P_i^g \leq \overline{P}_i^g && \text{(gen. limits)} \\
& \underline{Q}_i^g \leq Q_i^g \leq \overline{Q}_i^g && \text{(gen. limits)} \\
\forall (i,j)\in\mathcal{L}&: \nonumber\\
 & \theta_{ij} := \theta_i - \theta_j && \text{(angle deviation)} \\
& P_{i,j} = V_iV_j\left[G_{ij}\cos(\theta_{ij}) + B_{ij}\sin(\theta_{ij})\right] && \text{(branch flow)} \label{eqn:p-branch-flow}\\
& Q_{i,j} = V_iV_j\left[G_{ij}\sin(\theta_{ij}) - B_{ij}\cos(\theta_{ij})\right] && \text{(branch flow)} \label{eqn:q-branch-flow}\\
& P_{i,j}^\text{loss} + P_{j,i}^\text{loss} = P_{i,j} + P_{j,i} && \text{(branch losses)} \\
& Q_{i,j}^\text{loss} + Q_{j,i}^\text{loss} = Q_{i,j} + Q_{j,i} && \text{(branch losses)} \\
& |y_{ij}|^2\left[V_i^2 + V_j^2 - 2V_iV_j\cos(\theta_i - \theta_j)\right] \leq \overline{I}_{i,j}^2 && \text{(thermal limit)} \label{eqn:thermal-limit}\\
\textbf{HDEV Constraints}:\quad & \nonumber\\
\forall i\in\mathcal{N}&: \nonumber\\
& x_i(t) = \sum_{a\in\mathcal{X}_i(t)}\delta_a\sum_{h\in\mathcal{H}} \lambda_a^h && \text{(bus demand)} \\
& \underline{x}_i \leq x_i(t) \leq \overline{x}_i && \text{(station limit)} \\
\forall h\in\mathcal{H}, a\in\Arcs &: \nonumber\\
& \lambda_a^h \geq 0 && \text{(flow direction)} \\
\forall h\in\mathcal{H}, v\in\Vertices &: \nonumber\\
& \overline{\psi}^{h,\texttt{in}}, \overline{\psi}^{h,\texttt{out}} \sim \textbf{Travel Criteria Distribution} && {\begin{aligned} & (\text{user-specified} \\ & ~~~\text{arrival \& departure}) \end{aligned}} \\
& \qquad\qquad\qquad\qquad\text{s.t.}~ \sum_{v\in\Vertices} \overline{\psi}^{h,\texttt{in}}_v = \sum_{v\in\Vertices} \overline{\psi}^{h,\texttt{out}}_v && \nonumber \\
& \qquad\qquad\qquad\qquad\qquad \overline{\psi}^{h,\texttt{in}}_v, \overline{\psi}^{h,\texttt{out}}_v \geq 0 && \nonumber \\
& && \nonumber \\
& \sum_{a\in\mathcal{O}_v} \lambda_a^h - \sum_{a\in\mathcal{I}_v} \lambda_a^h = \overline{\psi}^{h,\texttt{in}}_v - \psi^{h,\texttt{out}}_v && \text{(flow balance)} \\
& \psi^{h,\texttt{out}}_v \geq 0&& \text{(departure flexibility)} \\
& \sum_{u\in\mathcal{U}_v} \psi^{h,\texttt{out}}_u = \overline{\psi}^{h,\texttt{out}}_v && \text{(departure flexibility)}
\end{align}
\end{problem}
\end{table*}